\documentclass[preprint]{elsarticle}

\usepackage{lineno,hyperref}

\journal{Energy Strategy Reviews}

\usepackage[utf8x]{inputenc}

\usepackage{subscript}

\usepackage{todonotes}

\usepackage{multirow}
\usepackage{tabularx}
\usepackage{rotating}
\usepackage{amsmath, amssymb}
\usepackage{adjustbox}
\usepackage{booktabs}
\usepackage{pifont}
\newcommand*\rot{\rotatebox{90}}
\newcommand*\OK{\ding{51}}

\usepackage{xcolor}
\hypersetup{
    colorlinks,
    linkcolor={!black},
    citecolor={!black},
    urlcolor={!black}
}

\begin{document}

\begin{frontmatter}

\title{The Open Energy Modelling Framework (oemof) - A new approach to facilitate open science in energy system modelling\tnoteref{t1}}
\tnotetext[t1]{This document is a collaborative effort.}


\author[znes,euf]{S.~Hilpert\corref{cor1}}
\ead{simon.hilpert@uni-flensburg.de}
\cortext[cor1]{Corresponding author, Phone: +49 (0) 461 805 3067}

\author[znes,euf,hsfl]{C. Kaldemeyer}

\author[rli]{U. Krien}

\author[znes,euf]{S. G\"{u}nther}

\author[znes,euf]{C. Wingenbach}

\author[rli]{G. Plessmann }

\address[znes]{Center for Sustainable Energy Systems (ZNES), Flensburg}
\address[euf]{Department of Energy and Environmental Management, Europa-Universität
Flensburg, Auf dem Campus 1, 24943 Flensburg, Germany}
\address[hsfl]{Department of Energy and Biotechnology, University of Applied Sciences Flensburg,
Kanzleistraße 91-93, 24943 Flensburg, Germany}
\address[rli]{Reiner Lemoine Institut gGmbH, Rudower Chaussee 12, 12489 Berlin, Germany}

\begin{abstract}

Energy system models have become indispensable tools for planning future energy systems by
providing insights into different development trajectories. However, sustainable systems with
high shares of renewable energy are characterized by growing cross-sectoral
interdependencies and decentralized structures. To capture important
properties of increasingly complex energy systems, sophisticated and
flexible modelling tools are needed. At the same time, open science is becoming
increasingly important in energy system modelling.
This paper presents the Open Energy Modelling Framework (oemof) as a novel
approach to energy system modelling, representation and analysis.
The framework provides a toolbox to construct comprehensive energy system models and
has been published open source under a free licence.
Through collaborative development based on open processes, the framework supports a maximum level of participation, transparency and
open science principles in energy system modelling.
Based on a generic graph-based description of energy systems, it is well-suited
to flexibly model complex cross-sectoral systems and incorporate various
modelling approaches.  This makes the framework a multi-purpose modelling
environment for modelling and analyzing different systems at scales ranging from urban
to transnational.

\end{abstract}

\begin{keyword}
decision support
\sep energy system modelling
\sep optimization
\sep collaborative development
\sep open science \linebreak\linebreak
\textcopyright { 2018, \href{
https://creativecommons.org/licenses/by-nc-sa/4.0/}
{CC BY-NC-SA 4.0}, 
\href{https://doi.org/10.1016/j.esr.2018.07.001}{
https://doi.org/10.1016/j.esr.2018.07.001}}

\end{keyword}

\end{frontmatter}


\section{Introduction}

The global transition process towards more sustainable and low-carbon energy systems requires
the development of alternative future trajectories for thorough scientific discussion.
Using these, decision processes on different levels e.g. in transnational
policy making or local energy planning can be supported.
However, future energy systems imply a rising complexity in technical, economic and socioeconomic
dimensions due to increasingly cross-sectoral and decentralized structures \citep{Pfenninger2014}.
Insights into such complex systems can be gained by applying computer-based modelling approaches
which create a quantitative basis for the above mentioned discussion and
decision processes.

Depending on the specific investigation and research question, a variety of model types can
be applied.
Such model types include power flow models for electricity transmission
network operation and planning, economic dispatch models for general capacity planning and
unit commitment models for power plant utilization \citep{Biggar2014, Harris2006, Wood2013, Kirschen2004}.
Applications range from large-scale transnational investigations using purely economic top-down
equilibrium models to detailed technical local infrastructure planning using bottom-up models based
on technology-specific data.
Moreover, many models can be adapted to integrate different sectors such as electricity, heat and
mobility to investigate cross-sectoral interdependencies.

Energy system models and derived results have often been heavily discussed among
different stakeholders and been criticized for not opening their internal logic and
underlying assumptions \citep{Pfenninger2017a, Pfenninger2017b, DeCarolis2012}.
As a result, in the last decade more scientists have opened their models and data
\citep{openmod2017, Pfenninger2018}. This process goes along with a general trend
to open science
in many other research fields. The rationale for open science includes improved
efficiency, scrutiny and reproducibility of results, re-usability of scientific work and
increased transparency of all scientific processes \cite{OECD2015}. As the European
Commission has recently started  to push open science in its research
programs \citep{EuropeanCommission2017}, the subject of openness has finally
moved into the public spotlight.

This paper presents the Open Energy Modelling Framework (oemof) as a novel approach
to foster open science in the field of energy modelling and analysis.
First, the idea of a single energy modelling framework is differentiated from other
approaches to delineate the scientific contribution in Section \ref{sec:scientific-contribution}.
The underlying  concept with its mathematical representation as well
as the framework architecture and its implementation are outlined in Section
\ref{sec:description}.
Building on this, the general  process of application development is described
in Section \ref{sec:applications} along with a selection of existing applications.
Finally, the general approach and its scientific contribution are summarized in
Section \ref{sec:conclusion}.

\section{Scientific contribution}
\label{sec:scientific-contribution}

To provide context, first a brief overview on relevant energy system
modelling software is provided.
Subsequently, the presented framework is compared to similar existing software
and its unique features are outlined.
For extensive reviews on this topic, please see\citet{Hall2016}, \citet{Conolly2010} and \citet{Pfenninger2014}.

\subsection{Overview of modelling landscape}

In the following, we distinguish between the three terms \emph{model},
\emph{model generator} and \emph{framework}. Models are concrete representations of
real world systems (e.g. with a specific regional focus and temporal resolution).
Such a representation may consist of multiple hard- or soft-linked sub-models
to answer clearly-defined research questions. Models can be built using model
generators that employ a certain analytical and mathematical
approach (e.g. by the use of predefined set of equations, represented technologies).
Finally, a framework can be understood as a structured toolbox including sub-frameworks
and model generators as well as specific models (e.g. wind feed-in models). In
addition this kind of a collection has other requirements for structures
and processes that guide the development process.

With respect to open science principles, a rough division into a line of closed
(\emph{1st generation}) and open (\emph{2nd generation}) models for
energy system analysis can be derived.

The \emph{1st generation} models and model generators have a long tradition and
are predominant in the academic energy system modelling field.  Among
the most widely used proprietary model generators is
TIMES/MARKAL \citep{Loulou2004}. Models of this family have been used
to answer research questions in the field of energy planning which is indicated by the
high number of references in academic literature \citep{Hall2016}. Similarly,
MESSAGE is a prominent model generator that has been used for the IIASA global
energy scenarios \citep{iiasa1542}. Besides this, the EnergyPLAN simulation
model has been applied in various research projects to analyse sector
integrated electricity, heat and transport systems \citep{Aalborg2017}.

The Balmorel model \citep{Ravn2001} can be seen as one of the first
\emph{2nd generation} energy system models.
It has been designed for power and heat dispatch modelling with optional
investment within the Baltic region and is written in GAMS.
Another early project is the model generator OSeMOSYS \citep{Howells2011} which is mainly
used for long-term integrated assessment and energy planning.
This project aims to facilitate modelling and education through a free software philosophy
and a simple, easy-to-learn interface.
Since then, various other projects with different purposes have been developed
(e.g. urbs \cite{urbs07}, PyPSA \cite{PyPSA}, calliope \cite{Pfenninger2015}).
Their focus covers the full range from power flow simulation to long-term investment models.
A list of open source models can be found on the website of the Openmod-Initiative \citep{openmod2017}.
While some of these projects are models for a specific region, others can be classified as
model generators.

\subsection{Comparison to other software}

Since the list of available of modelling software is extensive, the framework is
compared to similar existing tools. For this, major categories with single
characteristics are introduced. These encompass the general suitability for
\emph{open science}, the technical \emph{concept} and overall
modelling \emph{functionality}.

A requirement of \emph{open science} is the free availability of the
software itself.
Freely available software is software that is available without additional cost.
The usage of fee-based software creates barriers to
reproducibility, since the experiment can only be repeated if the respective licence
is procured. Moreover, an open licence enables users to distribute, understand and change the
source code and thus enhances transparency since model assumptions and internal logic
can be understood, changed and evaluated to determine their influence on the
results.
The issue of re-usability can also be addressed when software is published under an
open licence since other developers can re-use any part of the software.
Finally, collaborative software development allows for continuous improvements,
enables an easier detection of bugs and makes it possible to discuss new features
in a transparent manner.
Collaborative development in this context refers to joint work on the
software's code without mandatory institutional ties. This includes a
common road map, discussion of new features and changes and in general a high
level of communication among the developers. A central characteristic of this
definition is the transparency of all associated processes.

The \emph{concept} is defined by technical and structural characteristics.
Implementing the software in a high-level language lowers barriers to usage and
contribution. High-level programming languages are characterized by a strong
abstraction from computer's hardware, are easier to use and understand, may
include elements of natural language and make software development simpler.
The more external libraries available for a language, the easier the implementation of various tasks in the
modelling tool-chain. Further, interfaces to other languages can be used to extend capability.
A generic data model enables a separation
between the mere topological description and subsequent calculation (within
an optimization, for example).
Generic data models are data structures designed specifically to suit
the representation of data of a particular problem rather than to
store data of multiple different problems.
Graph-based representation of energy systems, for instance, can be used to
represent electricity systems as well as heating systems.
Providing the option to define the level of accuracy flexibly is an added value of
energy system modelling toolboxes. For example, it allows for user-defined
precision in representation of time in modelling an energy system
components by extending the libraries scope through user-defined components.
Another aspect of the concept is designing it for multiple purposes. This extends
the core functionality by other useful tools.
An example would be an energy system modelling toolbox that includes tools for generating
feed-in profiles of renewable energy sources.

\emph{Functionality}, in this case, is defined by concrete modelling capabilities
for model types such as economic dispatch, investment planning (also across
multiple periods, called multi-period investment planning), power flow calculation and
unit commitment. Furthermore, the general capability to model sector coupling
problems is a prerequisite for modelling multi-sectoral energy systems such as electricity, heat and transport.

To compare the framework to other tools, Table \ref{tab:software_comparison}
lists a selection of popular \emph{1st} and \emph{2nd} generation of modelling frameworks
and model generators. Though oemof shares certain characteristics with existing
software, the collaborative development, the generic data model and multi-model toolbox (framework) differentiates oemof.

\begin{table}[ht] \centering
   \begin{adjustbox}{max width=\textwidth}
    \begin{tabular}{@{} cl c c c c c c c c c c c c c c @{}}
        & & \multicolumn{3}{c}{Open science} & \multicolumn{4}{c}{Concept} & \multicolumn{6}{c}{Functionality} \\
        \cmidrule[1pt](r){3-5} \cmidrule[1pt](lr){6-9} \cmidrule[1pt](l){10-15}
\\
        & & \rot{Free of charge} & \rot{Open licence} & \rot{Collaborative development}
        &  \rot{High-level language} & \rot{Generic data model} & \rot{Flexible level of accuracy}
        & \rot{Multi purpose toolbox} & \rot{Economic dispatch} & \rot{Multi period investment planning} & \rot{Investment planning}
        & \rot{Power flow}  & \rot{Unit commitment}  & \rot{Sector coupling} & \rot{Source}\\
        \cmidrule[1pt]{2-16}
        & WASP IV             & \OK  &     &     &     &     &     &    & \OK  & \OK & \OK  &     &     & \OK & \cite{wasp4}\\
        & EnergyPlan v12      & \OK  &     &     &     &     &     &    & \OK  &     & \OK  &     &     & \OK & \cite{Aalborg2017}\\
        & MARKAL/TIMES        &      &     &     &     &     &     &    & \OK  & \OK & \OK  &     &     & \OK & \cite{Loulou2004}\\
        & MESSAGE-III         &      &     &     &     &     &     &    & \OK  & \OK & \OK  &     &     & \OK & \cite{iiasa1542}\\
        \cmidrule{2-16}
        & oemof v0.2         & \OK  & \OK & \OK & \OK & \OK & \OK & \OK& \OK  &     & \OK  &     & \OK & \OK & \cite{oemof014}\\
        \cmidrule{2-16}
        & urbs v0.7	      & \OK  & \OK &     & \OK &     &     &    &  \OK &     & \OK  &     & \OK & \OK & \cite{urbs07}\\
        & calliope v0.5.3     & \OK  & \OK &     & \OK &     &     &    &  \OK &     & \OK  &     & \OK & \OK & \cite{caliope053}\\
        & PyPSA v0.12          & \OK  & \OK &     & \OK &     & \OK &    &  \OK
&     & \OK  & \OK & \OK & \OK & \cite{PyPSA}\\
        & OSeMOSYS            & \OK  & \OK &     &  (\OK)     &     &     &
&  \OK & \OK & \OK  &     &     & \OK & \cite{Howells2011}\\
        \cmidrule[1pt]{2-16}
    \end{tabular}
    \end{adjustbox}
    \caption{A comparison of features of selected software tools that are
similar to oemof. Note that for \emph{OSeMOSYS} multiple implementations in
different languages exist.}
    \label{tab:software_comparison}
\end{table}

\subsection{Unique framework features}

A core feature of the framework is its \emph{collaborative development} with the
goal of \emph{community building}.
Many existing tools are not developed by a single institution. For example,
researchers are encouraged to develop and improve the source
code of MARKAL and other tools of the Energy Technology Systems Analysis
Program (ETSAP)\cite{ETSAP2017}. However, the review processes and decisions are
not transparent and valuable information may be lost in case of rejection of
input.
In contrast, oemof strives for an open process to encourage future improvements.
To align with open science principles the idea is to enable full transparency of
the development process and not only the final source code.
For that reason, the project follows a strict free software philosophy. In addition,
processes are  designed for community building, collaborative and transparent
source code development.

Another unique feature is the \emph{generic data model} which has emerged from
the collaboration of various researchers with different research interests and backgrounds.
This has led to the development of a framework with a common basis
(Section \ref{sec:underlying_concept}) consisting of a layer-structured set
of tools and sub-frameworks.
A generic graph-based basis allows to differentiate between the topology of an
energy system and its calculation based on a specific mathematical approach.
The oemof framework may be seen as a domain specific
language \cite{Booch2005} that represents arbitrary energy systems as a graph.
As a consequence, oemof can represent energy systems at
a high abstraction level as well as a detailed single power plant.

Generally, the framework serves as a \emph{multi-purpose toolbox} for energy system
modelling and has been designed to integrate a growing set of toolboxes in future.
Open source model generators like \emph{calliope} \citet{Pfenninger2015}
and the toolbox \emph{OseMOSYS} \citet{Howells2011} are designed
to build specific models of one model family or type by the use of
predefined sets of equations (e.g. bottom-up linear optimization based models).
Furthermore, some of the existing projects, such as \emph{PyPSA} \cite{PyPSA},
include several model generators for different purposes that may be combined.
In contrast to other tools, oemof encompasses model generator methods to generate specific economic
dispatch, investment and unit commitment models.  Beyond this, it
provides a structured set of tools to facilitate the modelling process. In its
current state, this set includes an optimization library (model generator) as well as
tools to simulate feed-in from renewable energy sources or local heat demand for
a specific region.

In summary, the underlying concept, the software architecture, the free
software philosophy and in particular the framing processes (e.g. open meetings,
open code review, open web-conferences, open platforms and open pull-requests)
enable  collaborative development and participation. These combined features
distinguish oemof from existing projects and constitute a basis for open science
in the  field of energy system modelling. Its academic value lies exactly in
this difference in terms of open science.

\section{Concept, architecture and implementation}
\label{sec:description}

To help in understanding the framework, its underlying concept,
architecture and specific implementation are outlined here.
First, a general mathematical representation of
energy systems is proposed which serves as a base for higher level
software architecture presented subsequently.
Finally, the specific implementation is described and justified.

\subsection{Underlying concept}
\label{sec:underlying_concept}
The main feature of the framework is the separation of an energy
system's topological description from its computation.
The representation may serve as a foundation to run graph-based algorithms
(to determine whether the graph is connected, for example) or to perform exploratory analyses.
Subsequently parameters of the system (or sub-systems) can be computed based
on concrete modelling methods. Due to this property, oemof can incorporate
other models and model generators with varying modelling approaches and
different programming languages.

To achieve this, a generic
concept which constitutes the foundation of all the oemof libraries has
been developed. In this concept, an energy system is represented as a network consisting
of nodes and edges connecting these.
Nodes $N$ are subdivided into buses $B$ and components $C$. When
representing an energy system, an additional constraint that buses are solely
connected to components and vice versa is imposed.
Components are meant to represent actual producers, consumers or processes of the energy
system while buses are meant to represent how these components are tied
together. Edges are used to represent the inputs and outputs of a component.

An energy system that is represented in such a way can be mathematically
described using concepts from graph theory by looking at it as a bipartite
graph $G$. The mathematical formulation of this graph
in its general form is given in equation \ref{eq:graph}. A more detailed
description of this concept with its theoretical foundation has been
published by Wingenbach \citet{Wingenbach2016}.

\begin{align}
\label{eq:graph}
G := (N, E)\\ \notag
N := \{B, C\}\\ \notag
E \subseteq  B \times C \cup C \times B\\ \notag
C^+ \subseteq   C\\ \notag
C^- \subseteq  C\\ \notag
T \subseteq   C
\end{align}

Components can be subdivided further into sources $C^+$, sinks $C^-$ and
transformers $T$:
\begin{enumerate}[1.]
 \item \emph{Transformers} have inflows and outflows. For example, a gas turbine
       consumes gas from a gas bus and feeds electrical energy into an
       electricity bus. The relation between inflow and outflow can be
       specified in the form of parameters, for example by specifying the
       transfer function or an efficiency factor.
 \item \emph{Sinks} only have inflows but no outflows. Sinks can represent consumers
       of which households would be an example.
 \item \emph{Sources} have outflows but no inflows. For example, wind energy or
       photovoltaic plants but also commodities can be modelled as sources.
\end{enumerate}

\begin{figure}[!ht]
 \centering
 \includegraphics{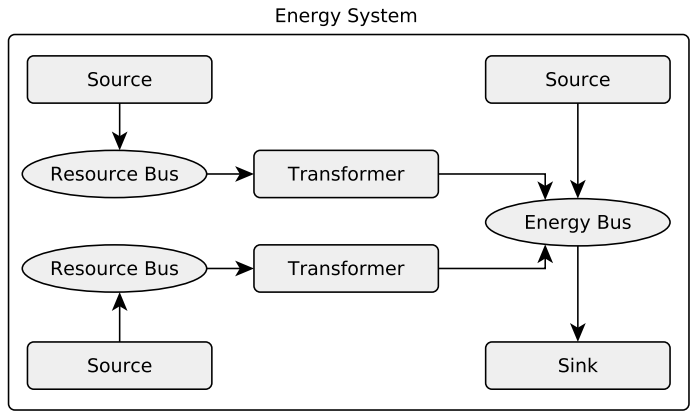}
 \caption{Schematic illustration of an energy system represented as an oemof
	  network.}
 \label{fig:abstract_oemof_network}
\end{figure}

A similar, purely mathematical
formulation of multi-commodity network flow optimization models for dynamic energy
management has been illustrated by Zeng \citet{Zeng2012}.
Furthermore, related structures of energy systems can also be found in different
energy models \citep{Howells2011, Fishbone1981, Loulou2004, Richter2011}.
These publications demonstrate that using a graph is an intuitive way of
representing an energy system. The main difference of our approach when compared to
existing ones is the identification (and its object-oriented implementation) of
a specific graph structure that may be used as a representation for all types of
energy systems. Every calculation based on a specific model will be derived
from this representation.
A graphical representation of how to describe an arbitrary energy system using
this network structure is shown schematically in Figure~\ref{fig:abstract_oemof_network}.

Based on the described concept, oemof provides basic components which can be
used directly while also facilitating the development of more specific
components built upon the basic ones.

\subsection{Mathematical description: The solph-library}
\label{sec:mathematical_description}

Currently, the solph-library can be used to create mixed-integer linear
optimization problems from a pre-defined set of components.
In order to model different elements of an energy system, several classes that may
represent real-world objects such as power plants or consumers
based on the described graph logic are provided.
Every class has associated objective expression terms,
optimization variables and constraints. Depending on the
object attributes set by the user these associated terms will be added to the model.

The objective function for a specific model consists of different
expressions depending on chosen components and their attributes.
Hence, only a
general description for main categories of distinct expression types is given
in this section. Generally, total costs for the simulated time horizon $T$
are minimized, whereas the expression term includes
time-dependent terms for all variables $w$ associated with an edge $(s,e)$ (i.e.
connecting a start node $s$ and an end node $e$) and for all variables $v$
associated with a node $n$. In addition, time-independent terms for node and
edge weights may be added.

\begin{align}
\text{min:~}
  & \overbrace{\sum_{t \in T} \Bigl(
    \sum_{(s,e) \in E}
      \sum_{i \in I_1}
	c_{(s,e)}^{i}(t) \cdot w^i_{(s,e)}(t) \cdot \tau
  \Bigr)}^{\text{\tiny{t-dependent expression for edge weights}}} \\ \notag
  &+ \overbrace{
    \sum_{(s,e) \in E} \sum_{i \in I_2} c_{(s,e)}^{i} \cdot w^{i}_{(s,e)}}^{
      \text{\tiny{expression for edge weights}}} \\ \notag
  &+ \overbrace{\sum_{t \in T} \Bigl(
    \sum_{n \in N} \sum_{i \in I_3} c_{n}^{i}(t) \cdot v^{i}_{n}(t) \cdot \tau
    \Bigr)}^{\text{\tiny{t-dependent expression for node weights}}}\\ \notag
  &+ \overbrace{\sum_{n \in N} \sum_{i \in I_4} c_{n}^{i} \cdot v^{i}_{n}}^{
    \text{\tiny{expression for node weights}}}
\end{align}

The parameter $c$ may be interpreted as a specific cost and the time-increment
$\tau$ is determined by the temporal resolution.
Sets $I_1$ to $I_4$ stand for the possibility of multiple costs and weights for one
edge or node. Domains $D$ of variables $w$ and $v$ can either be positive reals,
positive integers or binary, as a special sub-type of integer.

\begin{align}
 D = \{\mathbb{R}^+, \mathbb{Z}^+, \{0,1\}\}
\end{align}

Generally, all variables are bounded by lower and upper bounds which are set
based on the class attributes of the modelled components.

\begin{align}
 & 0 \leq w_{(s,e)}^i(t) \leq \overline{w}^i_{(s,e)}(t)
  & \qquad \forall i \in I_1,~ \forall (s,e) \in E,~ \forall t \in T\\
 & 0 \leq w_{(s,e)}^i \leq \overline{w}^i_{(s,e)}
  & \qquad \forall i \in I_2,~ \forall (s,e) \in E\\
  & 0 \leq v_{n}^i(t) \leq \overline{v}^i_{n}(t),
  & \qquad \forall i \in I_3,~ \forall n \in N, \forall t \in T\\
 & 0 \leq v_{n}^i \leq \overline{v}^i_{n},
  & \qquad \forall i \in I_4,~ \forall n \in N
\end{align}

The library consists of a large set of constraints that are
documented extensively in the latest online documentation of the software. In
addition, the library is being continuously improved. Therefore, possible constraints
are subject to changes and depend on the version of the library. For these
two reasons the constraints are not outlined in detail. Instead, a general form
of constraint which all specific component constraints must follow is given in
equation \ref{eq:general_constraint_form}.

\begin{align}
 \sum_{k \in P_n} \sum_{j \in J_1} a^j_{(k,n)} \cdot w^j_{(k,n)} +
 \sum_{k \in S_n} \sum_{j\in J_2} a^j_{(n,k)} \cdot w^j_{(n,k)} +
 \sum_{j \in J_3} a^j_n \cdot v^j_{n} + M \leq 0 & \qquad \forall n \in
N \label{eq:general_constraint_form}
\end{align}

The important characteristic of this constraint is the reduced possibility
space of related variables inside one specific constraint. Defined from the
perspective of a node $n$, only variables $w$ and $v$
associated with an edge from one of its predecessors $P_n$ to node $n$, an edge
from node $n$ to one of its successors $S_n$, or the node itself may appear. In
this context, the parameter $a$ may be interpreted as an efficiency, for example. The
sets $J_1$ to $J_3$ stand for the possibility of multiple parameters and weights
for one edge or node.

\subsection{Project architecture}
\label{sec:project_architecture}
\begin{figure}[ht]
 \centering
 \includegraphics[width=0.85\textwidth]{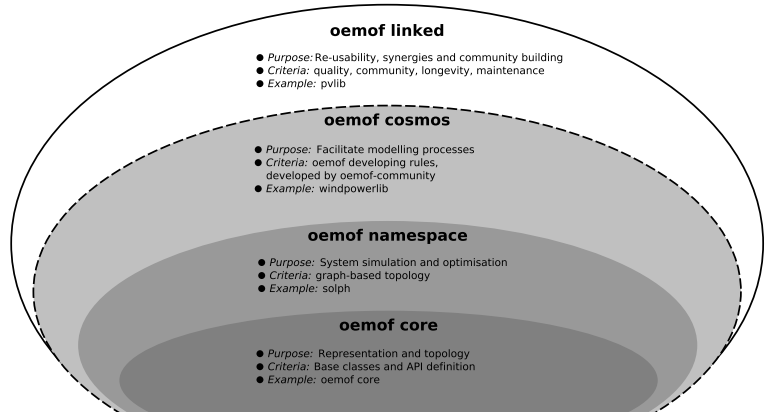}
 \caption{Layer structure of the oemof project architecture.}
 \label{fig:oemof_cosmos}
\end{figure}

The project tries to accommodate energy system modellers with a large set of
functionalities they typically need. To achieve this, the project and its
development process follow an architecture that groups the content of the
framework into functional and organizational units. This architecture consists
of the four layers
depicted in Figure~\ref{fig:oemof_cosmos}. These four layers are used to
categorize the libraries associated with the oemof project according to their
dependencies and commonalities.

The framework itself and its underlying concept are implemented using an
object-oriented approach in the high-level programming language Python
and they are published under the GNU GPL3 licence. Python libraries are
called packages; the main component of the framework is the oemof package.
This package covers the first layer completely and the second layer partially.
The layers beyond the first differentiate how closely libraries are associated
with the oemof package and its developers in terms of organizational ties as
well as technical dependencies.

\begin{enumerate}[1.]
  \item
    At the \emph{core} layer a generic graph structure is implemented via core
    classes. These classes are used to instantiate the objects comprising
    an energy system graph and define how an energy system is described. In addition,
    the basic application programming interface (API) is defined, through which
    the core objects and their properties representing the graph are
    accessed. The entire
    \emph{core} layer is kept free from energy system-specific logic in order
    to accommodate a broad spectrum of modelling approaches. Additionally, it
    allows decoupling of the energy system's representation from how it is
    modelled. The intended use of the core objects in layers above the \emph{core}
    layer is communicated via carefully chosen naming and explicit documentation.

  \item
    The \emph{namespace} layer contains associated libraries that share the
    same basic system formulations, i.e. libraries modelling energy systems
    as graphs described in terms of objects from the \emph{core} layer.
    They depend on the basic API by either directly using \emph{core}
    classes or adding functionality on top of them via inheritance.
    That way, different modelling approaches can be used on energy systems
    described in a uniform way, namely as energy system graphs consisting of
    instances of \emph{core} classes or of classes inherited from them.
    Possible modelling methods can model energy systems
    with respect to cost, power-flow or any other kind of simulation or
    optimization goal. Currently the \emph{oemof.solph} can be used to
    generate linear (mixed-integer) optimization problems from an energy
    system representation based on core objects.
    However, the graph structure is capable of accommodating other concepts
    such as evolutionary optimization or agent-based modelling.
  \item
    The \emph{oemof cosmos} layer contains libraries from the field of energy
system
    modelling that are associated with oemof in an organisational way without
    sharing the basic API. These libraries, while still part of the oemof project,
    are not developed as part of the same package and may thus be used, reviewed
    and developed by third-party modellers and experts as well. However, as they
    are developed as part of the framework, they follow the common development
    rules (Section \ref{subsec:documentation}). As most modellers are not
    primarily programmers, sharing the same development, structure and
    documentation rules can help in learning how to use the libraries.
    One example of such a library is the windpowerlib \citep{Haas2017}, a
    library generating feed-in time series of wind energy turbines from
    meteorological data.
  \item
    Open source does not necessarily lead to cooperation \cite{Pfenninger2018}.
    To facilitate cooperation, the \emph{oemof linked} layer contains
    existing community libraries. These libraries are written in
    Python but do not necessarily share the same rules. However, in order to be
    considered associated, they should meet general standards for quality,
    code development, longevity, maintenance and community structure.
    One example of such a library is the pvlib \citep{Holmgren2017}, which
is a
    library developed independently from oemof and which will be integrated
    into the framework via interfaces in the feedinlib. The process of
    developing these interfaces has already lead to code contributions towards
    pvlib, instead of
    the creation of a parallel, competing solution.
\end{enumerate}

\subsection{Implementation}
\label{subsec:implementatio}
\begin{figure}[h!]
 \centering
 \includegraphics[width=0.8\textwidth]{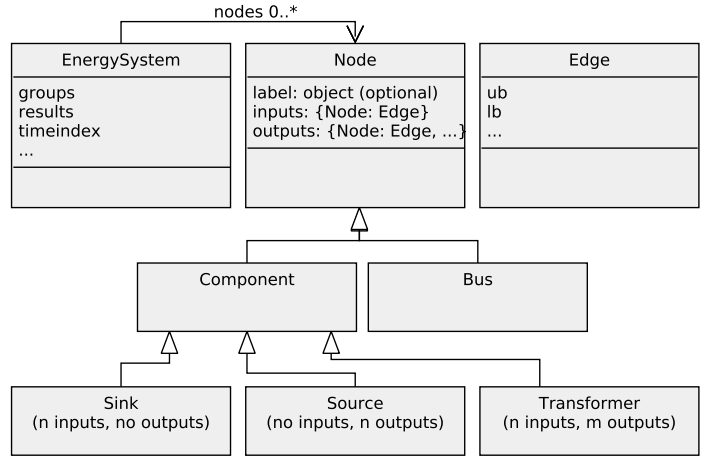}
 \caption{UML class diagram of oemof core classes.}
 \label{fig:oemof_classes}
\end{figure}

The graph concept has been implemented at
the \emph{core} layer in the form of a class hierarchy which is sketched
in Figure~\ref{fig:oemof_classes}. The root elements of this class hierarchy
are \emph{Node}, \emph{Edge} and \emph{EnergySystem}.
\emph{Node} is the abstract base class for \emph{Bus} and
\emph{Component}, which are used to represent nodes in the bipartite graph
representing the energy system. Further, components
are subdivided into \emph{Source}, \emph{Sink} and \emph{Transformer} classes
depending on how they are connected to \emph{Bus} objects.
Objects of the class \emph{Edge} represent the directed edge
between two nodes, i.e. the connection between a \emph{Bus} and a
\emph{Component} object. The class \emph{EnergySystem}
serves as a container  for nodes and may hold additional
information about the energy system.

All basic energy system components such as energy demands, (renewable) energy
sources and transformers between different energy buses can be
modelled by means of these basic classes. Additional components that introduce
new features can be added via inheritance. If sub-classing is not
suitable, new classes can be created and used together with the core classes.
As an example, the \emph{solph} library introduces a storage class with
different individual parameters.

A demonstration of separation of the description
of the energy system from its computation can be seen through the introduction of
the \emph{Edge}
class, which is separate from the \emph{Node} class hierarchy. Objects of this
class hold information about the flow between two nodes, such as maximum available
transfer capacities of power line flows or whether the amount of a certain flow
is fixed and if so, its value. As
evidence of the generic flexibility, objects of this class are used
in the \emph{solph} library to build inter-temporal constraints for different
kinds of energy system optimization problems such as combined heat and power
modelling or unit commitment.

The \emph{EnergySystem} class serves as a container for the aforementioned
elements and provides the possibility of adding extra information such as grouping
structures or optimization parameters. Additionally, it provides interfaces to save and
restore the energy systems instance and to process results. This allows for an
intuitive handling of energy systems by treating them as their own entity.

An implementation using the high level programming language Python
has the advantage of a rich set of external libraries usable for scientific
computing. Oemof itself makes heavy use of external modules for
optimization problems (pyomo \cite{Hart2012}) and data handling
(pandas \cite{McKinney2012}).

\subsection{Documentation, collaboration and testing}
\label{subsec:documentation}
`A critical part of any piece of software is the documentation' has already been
stated by Greenhall and Christie \citep{Greenhall2012}.
This is of particular importance for open source projects
with many users and a changing developer base. With the objective of thorough
documentation in all stages and formulation of general nomenclature, a
documentation strategy on four different levels similar to that of Howells \citet{Howells2011}
is followed:

\begin{enumerate}[1.]
    \item \emph{Comments inside the code} are used to explain non-intuitive
lines of source code to new developers and interested users at the lowest level.
    \item \emph{Docstrings}
located inside the source code describe the API, i.e. how to use the various
classes, methods, and functions.
    \item \emph{Higher level descriptions}
provide the user with additional information about the possible
interactions between different libraries or application-specific usage
information. These manuals are located inside the repository and are
therefore shipped with the source code.
    \item \emph{Examples} provide an additional source of documentation that
    is particularly useful to new users and developers.
\end{enumerate}

Keeping such detailed documentation consistent and up-to-date across
continuous releases comes at the expense of a high maintenance effort.
Nevertheless, it is of special importance; the oemof
documentation is the place to find information on the formulas
used within an oemof-based model. Up-to-date, consistent documentation that tracks changes in a timely fashion is essential
if external users want to understand the internal logic of a model,
especially in scientific applications.
The upside is that documentation adhering to these principles acts as a
citable source of information, reducing the amount of redundant
information that must be sourced and digested in order to understand a
model. This in turn increases transparency and comparability.

As oemof is an open-source community project, a common platform for
collaboration is needed. Similar to \citet{Greenhall2012}
as well as other
open-source energy modelling projects, oemof uses GitHub for collaboration, code
hosting and bug tracking, which allows for easy copying and forking of the
project. To lower entry barriers for new developers, hierarchies for all
processes are kept as flat as possible. We have found that this can create a sense of
belonging for collaborating developers which increases participation.
GitHub is based on the version control system \emph{git} and code can be
developed in parallel on different branches. In order to ensure an effective
branching strategy and release management, a well-established  git workflow
model \citep{VanDriessen2010} is set as the standard for all developers.
Contributions to the code base are managed through pull requests, which allow
for an open review of potential changes. Further,
code changes are checked for conflicts before being merged back into
the development branch by the developer in charge of the affected
library.

In order to test oemof's functionality in case of changes to multiple
parts of the code base, \emph{unit tests} are employed. During the testing
process, all integrated application examples are run and the created results are
checked against stored historical results. Only if all examples run without errors is a
pull request merged back into the development branch. This procedure ensures
the functionality even if major changes to the code base are applied from one
release version to another.

\section{Usage: Applications}
\label{sec:applications}

The framework is not designed to be a standalone executable.
Instead, the oemof libraries are meant to be used in combination to build
energy system models. In the following we will refer to such models as
oemof-\emph{applications}.

\subsection{Application development}

\begin{figure}[ht]
 \centering
 \includegraphics{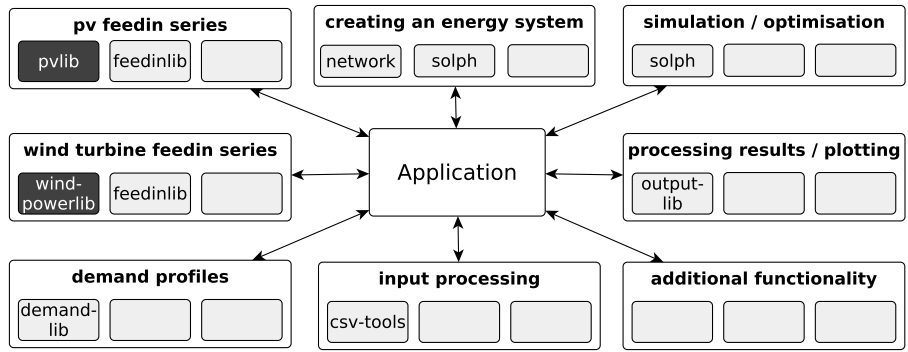}
 \caption{Building an application based on libraries of the oemof cosmos and
external libraries (dark grey).}
 \label{fig:oemof_app_building}
\end{figure}

Applications can be developed by the use of one or more framework libraries
depending on the scope and purpose. Figure~\ref{fig:oemof_app_building} illustrates
an example process of building an application. Modelling can thus range
from a few plain steps in a standalone Python executable to complex
procedures bundled in a new Python library based on oemof.
Due to the modular concept, specific functionalities of oemof libraries can
be substituted easily depending on the modelling task. This provides a high degree
of freedom for developers, which is particularly relevant in scientific working
environments with spatially distributed contributors and fast evolving research questions.

Depending on the problem, input data can be created by means of libraries such
as the \emph{feedinlib} or \emph{demandlib} library. A standardized result
processing library (\emph{outputlib}) provides all optimization results in convenient
data structures that are ready for exports to different formats, detailed analyses
and plotting. Although this feature might appear trivial, it is one major advantage
over other heterogeneous optimization tool-chains that require switching
between
tools e.g. GAMS for the modelling and a spreadsheet-based solution for result processing.

However, in considering the modelling workflow for the oemof namespace layer,
all applications have some major steps in common and include all
required data pre- and post-processing.
First, an empty energy system object is created.
This object acts as a container for the nodes and carries
information such as the time resolution. The energy system object may also hold
different variants of the system representing different scenarios.
Additionally, methods to handle nodes are provided. The next step is the
instantiation of nodes and flows of the modelled energy system which are
added to the existing energy system instance (population of energy system).
Subsequently, the results of the energy system can be computed by
simulating or optimizing the system. Finally, results can be processed
with the output library of oemof. The \emph{oemof-outputlib} makes it easy to
get different views of the results and plots based on a uniform output data format.

\subsection{Example application workflow: System optimization}
One common use case for a modelling process that utilizes different toolboxes
is the optimization of energy systems. In this process the \emph{solph} library can be
used in combination with existing input and output data libraries. First,
feed-in data for renewable power plants and electricity demand profiles can be generated
within the \emph{feedinlib} and \emph{demandlib} libraries. Subsequently, the data
are used as exogenous parameters within the \emph{solph} library before the
optimization results are processed within the \emph{outputlib}.

The \emph{solph} library allows the creation of mixed-integer linear models.
As a common requirement, an energy system graph has to be created with classes from the \emph{core} layer, respective \emph{solph} subclasses from the \emph{namespace} layer or
a mix of both. The energy system serves as a container that holds all nodes and
general information such as the temporal resolution of the optimization problem.
Since an oemof optimization model inherits from a model of the \emph{pyomo} package,
the full functionality of this package can be leveraged. Depending on the
experience and modelling task, three different ways exist to create an optimization
problem based on an oemof energy system instance.

\begin{enumerate}
 \item In the most common and easiest use case, the energy system describes
 a graph with flows on its edges by combining basic components and
 buses. The optimization model for this use case is automatically created by a
 logic that transfers the graph (connections between buses and components
 and their attributes) into respective constraints,  e.g. commodity
 balance equations or inequalities for  lower and upper flow bounds.
 When using this way of modelling, all models are derived by the object
 parametrisation and no mathematical definitions like sets, variables or
 parameters have to be implemented.
\item In the second use case, basic energy systems can be adapted by defining
 additional constraints on top of the aforementioned graph logic. Since this logic
 is consistent throughout, entry barriers for new users are lowered. As
 one example, an annual limit on a commodity flow can be implemented easily by a
 definition of (in)equations applied to a set of flows.
\item In the third use case, custom components can be added to a model. This
 is possible by subclassing from core components  or by creating one's own components from
 scratch.  As mentioned before, the full functionality of the \emph{pyomo} package
 can be utilized to model complex internal relations of components with
 numerous constraints, specific sets and  different variable domains. Such a
 component needs to provide input/output slots that may be connected with
 flows of graph.
\end{enumerate}

All use cases can be applied separately or combined within one model.
The model type itself, e.g. an economic dispatch, investment or unit commitment model,
is determined by its parametrisation. This allows for maximum flexibility, as one can
quickly change the model, from economic dispatch to investment, for example, by exchanging single
components, say a storage with fixed capacity (parameter) by one with variable
cost-determined capacity (decision variable), for example.
In a similar way, complete sub-graphs can be exchanged quickly by connecting or
disconnecting them from a main problem.

\subsection{Existing applications}

\begin{figure}[ht]
 \centering
 \includegraphics{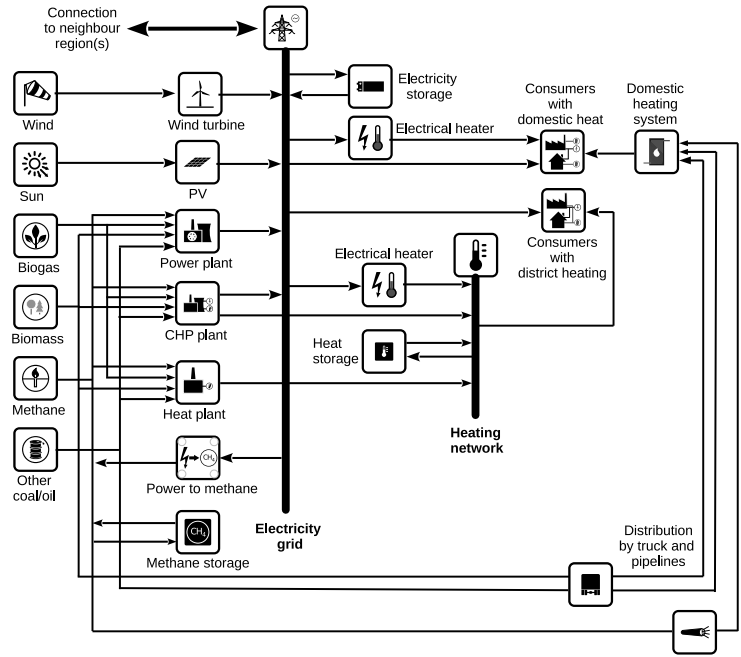}
 \caption{Representation of a complex energy system within an oemof application.}
 \label{fig:oemof_complex_application}
\end{figure}

The framework has already been used to build comprehensive applications
for different research projects (see \citep{Moller2016, Mueller2017, vernetzen2016, Wingenbach2016a, Kaldemeyer2016, Arnold2017}).
In addition, oemof is also used actively in teaching by some institutions in order to
gain insight into complex energy systems.
An example for such a system modelled as an oemof application is illustrated
in Figure~\ref{fig:oemof_complex_application}.
In the following, selected oemof applications are described to illustrate the
broad range of applications.
These distinguish themselves by technologies considered, demand sectors modelled,
regional representation, the time horizon of the analysis, the modelling methodology
to represent technological characteristics and perhaps a market representation.

The renewable energy pathways simulation system
(\emph{renpassG!S}) \citep{ZNES2016a} is a bottom-up fundamental Western European
electricity market model. Future scenarios of the power plant dispatch and price
formation in Germany and its interconnected neighbouring countries can be modelled
based on operational and marginal costs and the assumption of an inelastic
electricity demand.
Based on \emph{renpassG!S}, a spin-off model that is
adapted to the requirements of the Middle East and North Africa (MENA) region was created.
In this application the \emph{solph} library was used with a restriction to purely
linear equations.
Both applications use a standardized interface to csv-files for the
\emph{solph} library that was created to simplify the usability for users with no programming
experience.

The \emph{openMod.sh} application is a flexible software tool that is strongly based on
oemof's underlying concept \cite{Wingenbach2016a}. This model has been applied in
participative workshops for the development of regional climate protection scenarios.
The combination of a graphical browser-based user interface combined with an open-source
modelling approach enhances the modeller-decision-maker interface. The extension
of the underlying concept to a database concept shows that this concept may
not only be used for the computation of systems but also for
their representation in a relational database. Due to the open
licence and the high-level language, oemof applications can be set up
on public servers with little effort and without legal barriers.

An example for the flexible extension of oemof at the application level is the Heating System Optimization Tool (\emph{HESYSOPT}) \citep{Hilpert2016}.
In this application, detailed heating system components are modelled with mixed integer
linear programming techniques that are based on oemof.solph functionalities.
Using the underlying \emph{pyomo} library \emph{solph} provides an interface to add
new components within the application. After a review, such components can be integrated
into \emph{solph} to be available for the entire community.

As a fourth example, \emph{reegis\textsuperscript{hp}} \citep{RLI2016}
models heat and power systems on a local scale. The objective is to evaluate
district heating and combined heat and power technologies in energy systems
with a high share of renewable resources from an environmental and economical
perspective. The local system is connected in terms of electricity to a national
model based on the idea of the model \emph{renpassG!S} \citep{ZNES2016a} which is
extended to include the heating sector. This application uses oemof's
\emph{windpowerlib}, \emph{feedinlib} and \emph{demandlib} to provide the input
data for the model. Further, the \emph{solph} library is used to create a
large-scale linear model and a detailed mixed integer linear problem.
This example demonstrates how models of different scale may be combined in one
application.


These applications illustrate the flexibility of oemof and the extent of
the potential user group, not only with regard to the content, but also
concerning the level of involvement. It is possible to build a full-scale
energy system model adapted to the user's needs by just employing existing
functionalities. Moreover, different models can be combined and adapted
with little effort to create tools for specific purposes.
This enables users to answer challenging research questions
within a single framework.

\section{Conclusion}
\label{sec:conclusion}

The paper presents the Open Energy Modelling Framework (oemof)
as a contribution to the scientific modelling community. With a collaborative
and open development process, it is designed for transparency and
participation. Complementary to its technical features, the project
constitutes a novelty in energy system modelling and aims to facilitate open
science in this research field.

One central feature of oemof is the generic graph-based foundation which has
been implemented using an object-oriented programming methods in the high-level language Python. The cross-institutional collaborative development of the framework
has started a process towards this common and generic structure.
This concept highlights the distinction between the description of an
energy system
with its components and subsequent computations based on combining an intuitive
description with a specific mathematical approach.
It lays the foundation for a universal representation of
multi-sectoral energy systems at different scales.
Another important feature is its strict open-source and non-proprietary philosophy.
This philosophy, the underlying concept and the
extensive documentation allow new developers to adapt or extend the framework
easily and leverage features of other scientific Python libraries. With these
properties, the project is suitable for new developers and users and thereby supports
a continuous development process.

The framework has been successfully applied in different research
projects at several institutions.
Existing oemof applications include electricity market models,
detailed technical unit commitment models for district
heating systems and sector coupled regional energy system models.
Energy systems ranging from distributed or urban ones up to a
national scale may be modelled, making the framework a multi-purpose modelling
environment for strategic energy analysis and planning.

Although it takes effort for new users to learn to build an oemof-based application,
we think there are good reasons to choose oemof. Firstly, the flexibility
in application development allows adjustments along with changing
research objectives and may thus avoid lock-in effects. This seems to be
particularly relevant for project-based research. Secondly, the community
character of the oemof-project is another important factor. The possibility for
active participation in development and decision processes allows users to be part of
a community. We argue that this can create a sense of belonging, a value that goes beyond the technical features of the
software.

\section*{Acknowledgements}

The author group of this paper is a subset of the oemof developer
group. Since oemof is a collaborative project, all developers can be found
on the contributors page of each repository on GitHub.
The project was initially created by the following institutions:

\begin{itemize}
 \item Europa-Universität Flensburg (EUF)
 \item Flensburg University of Applied Sciences (HFL)
 \item Otto von Guericke University Magdeburg (OVGU)
 \item Reiner Lemoine Institut Berlin (RLI)
\end{itemize}

Special thanks and acknowledgement is owed to all people who have supported the project since it began.
In particular we want to thank Birgit Schachler (RLI), Caroline Möller (RLI),
Martin Söthe (ZNES), Wolf-Dieter Bunke (ZNES) for their individual
contributions.


\end{document}